# The "parent" state in kagome metals and superconductors: Chiral-nematic Fermi liquid state

Zihao Huang[1,2,3], Zhan Wang[1], Hengxing Tan[4], Zhen Zhao[1,2], Chengmin Shen[1,2], Haitao Yang[1,2], Bent Weber[3], Binghai Yan[4], Hui Chen[1,2*], Ziqiang Wang[5*], Hong-Jun Gao[1,2*]

[1] *Beijing National Center for Condensed Matter Physics and Institute of Physics, Chinese Academy of Sciences, Beijing 100190, PR China*

[2] *School of Physical Sciences, University of Chinese Academy of Sciences, Beijing 100190, PR China*

[3] *School of Physical and Mathematical Sciences, Nanyang Technological University, Singapore 637371, Singapore*

[4] *Department of Condensed Matter Physics, Weizmann Institute of Science, Rehovot, Israel*

[5] *Department of Physics, Boston College, Chestnut Hill, MA, USA*

[*]Correspondence authors, Email address: hjgao@iphy.ac.cn, wangzi@bc.edu, hchenn04@iphy.ac.cn

The kagome metals and superconductors hosting rich correlated and topological electronic states have captivated quantum materials research. These states are triggered by an unconventional chiral charge density wave (CDW) wherein a chiral superconductivity emerges at low temperatures, yet the origin of this chiral CDW order, the "parent" state, is unresolved. Here, we report the discovery of a parent chiral-nematic Fermi liquid state in kagome metals and superconductors. We use spectroscopic-imaging scanning tunneling microscopy to study Ti-doped $CsV_3Sb_5$ where the CDW is suppressed, and find that multiorbital Fermi surfaces break all mirror reflections and exhibit handedness. We observe the chiral low-energy quasiparticle dispersions, providing direct evidence for a chiral-nematic electronic structure. We further observe the direct transition from the parent state to a chiral-nematic superconducting state. Moreover, in the samples with chiral CDW, we also observe the residual chiral-nematic QPI features, demonstrating that the CDW-triggered exotic states descend from the chiral-nematic Fermi liquid. Our findings not only provide a plausible chiral-nematic parent state, but also establish a new conceptual framework for exploring the emergence and consequences of such exotic quantum phases.

In condensed matter and quantum materials physics, correlated electron states without symmetry breaking are described by the Landau Fermi liquid theory, where electron-electron interactions dress the electrons into low-energy quasiparticles[1] with renormalized energy-momentum dispersion near the symmetric Fermi surface (FS). In the Landau paradigm, symmetry-breaking correlated electronic states can emerge through the FS instabilities. For instance, nested FSs can be unstable toward translation symmetry breaking, giving rise to density waves such as charge and spin density waves[2-5] that reconstruct or fold the FS and quasiparticle dispersion. Interactions can also cause the FS to be unstable without disrupting translation symmetry, but toward point group symmetry breaking. FS deformation can arise from the Pomeranchuk instability[6-11], leading to a nematic Fermi liquid state with rotational symmetry breaking FS.

The kagome superconductors $A$V$_3$Sb$_5$ ($A$ = K, Rb, Cs) have garnered significant interest due to the recent discovery of landscape of correlated and topological electronic states triggered by a charge density wave (CDW) transition[12-24]. The CDW state is highly unconventional and breaks rotational symmetry in addition to translation symmetry. Remarkably, the CDW breaks all reflection symmetries about the mirror plane perpendicular to the kagome plane and thus exhibits a handedness or chirality[19,20,24]. Moreover, such a chiral CDW exhibits evidence for time-reversal symmetry breaking despite the absence of spin-related magnetism[19,20,24], which has led to the proposal of electron loop-current order[25]. At low temperature, the kagome metals transition from chiral CDW into an exotic superconducting state with evidence for pair density waves and chiral pair density modulations and chiral superconductivity[14,18,26,27]. The chiral CDW with intertwined symmetry breakings raises fundamental questions about (1) the origin of these extraordinary quantum states, the "parent" state, on the kagome lattice, and (2) whether the parent state is a symmetric Fermi liquid or already breaks point group symmetries. Addressing these issues will provide valuable insights for understanding the mechanism of the chiral CDW and the descendant phases, yet has been a challenge in quantum materials research.

Here, we report the discovery of a "parent" chiral-nematic Fermi liquid state by studying Ti-doped kagome metal CsV$_{3-x}$Ti$_x$Sb$_5$, where the CDW state is progressively suppressed at large Ti concertation $x$ and enables spectroscopic probes of the parent state at low temperatures. Using spectroscopic scanning tunneling microscopy, we image the coherent quasiparticle interference (QPI) pattern. Surprisingly, we observe robust low-energy QPI patterns in Ti-doped CsV$_3$Sb$_5$ that break all reflection symmetries about the mirror planes perpendicular to the kagome lattice and exhibit well defined handedness. The energy evolution of the chiral QPI patterns reveals the remarkable rotation and mirror symmetry breaking chiral-nematic FSs (such FSs are schematically illustrated in Fig. 1a) and quasiparticle dispersions involving both Sb *p* and V *d* orbitals. The kagome metal $CsV_{3-x}Ti_xSb_5$ in the absence of CDW order is thus identified as an unprecedented *multiorbital chiral-nematic Fermi liquid*. Intriguingly, the chiral-nematic deformation of the FSs can be captured by a novel *d*-wave quadrupolar distortion of the $C_6$ symmetric Fermi liquid due to intra-unit-cell sublattice charge order on the kagome lattice. Moreover, in lightly Ti-doped samples at small *x,* we observe restructured chiral QPI patterns due to the scattering of the residual quasiparticles in the CDW order state with broken translation symmetry. These observations point to the existence of a parent chiral-nematic Fermi liquid state in kagome metals, which genetically engenders chirality on the cascade of exotic quantum phenomena emerging upon lowering temperatures, either through an intervening chiral CDW phase or directly into the chiral superconducting state.

**Orbital resolved chiral QPI in the absence of CDW order**

The $CsV_3Sb_5$ single crystal has a layered structure with a stacking sequence of Cs-Sb2-VSb1-Sb2-Cs layers (Fig. 1b inset) and hexagonal symmetry (space group No.191, P 6/*mmm*). The kagome lattice of V atoms is coordinated by Sb1 atoms at the hexagon centers, forming a simple hexagonal lattice[37]. It has been shown that $CsV_3Sb_5$ can be doped by Ti that substitutes V sites within the kagome layer and create $CsV_{3-x}Ti_xSb_5$, where *x* is the Ti concentration[28]. To probe the electronic structure and reveal the electronic state without translation symmetry breaking, we first study $CsV_{3-x}Ti_xSb_5$ at $x = 0.15$, where long-range ordered CDWs are absent (Fig. 1b)[22, 28]. In the absence of other symmetry breaking, this state would be a symmetric Fermi liquid with $C_6$ symmetric FSs and quasiparticle excitations as predicted by the DFT calculations (Fig. 1c). The STM topography (Fig. 1d) obtained over a large-area on the Sb-terminated surface and the corresponding Fourier transform (FT) (Fig. 1e) show that the long-range $2a_0 \times 2a_0$ and $4a_0 \times 1a_0$ charge modulations, trademarks of the multiple CDWs in $CsV_3Sb_5$, are indeed absent at $x = 0.15$. The suppression of charge order enables direct investigation of the translation-invariant low-energy electronic structure via orbital resolved QPI measurements[22].

The spatial map of the tunneling conductance d*I*/d*V*(***r***, *V*) acquired at bias $V = -10$ mV over the same area reveals interference patterns (Fig. 1f). The corresponding FT d*I*/d*V*(***q***, -10 mV) resolves several well-defined scattering features, from which three sections of the prominent wave vectors, $\boldsymbol{q}_1$, $\boldsymbol{q}_2$, and $\boldsymbol{q}_3$ are identified (Fig. 1g). The QPI vectors correspond to scattering processes at the constant-energy ($\omega$) contours (CEC), described as $\boldsymbol{q}(\omega) = \boldsymbol{k}_f(\omega) - \boldsymbol{k}_i(\omega)$. By comparing with the calculated CEC (Fig. 1c) at $k_z = \pi$ near the Fermi level[22], we attribute $\boldsymbol{q}_1$ (Fig. 1g) to intra-pocket scattering within the Sb $p_z$ orbital-derived pocket (blue circular pocket $\alpha$ in Fig. 1c), while $\boldsymbol{q}_2$ and $\boldsymbol{q}_3$ (Fig. 1g) originate from inter-pocket scattering between in-plane V $d$ orbital-derived pockets (red triangular pockets $\delta$ in Fig. 1c). Note that consistent with previous results, the out-of-plane V $d$ orbital-derived FSs (green lines in Fig. 1c) are invisible (Fig. 1g) in the coherent QPI patterns[22].

## Observation of chiral-nematic Fermi surface

If the electronic state in the absence of CDW were a symmetric Fermi liquid, the QPI patterns resulting from Fig. 1c would be symmetric under 6-fold rotation ($C_3$ plus inversion). To check this, we compare the QPI patterns marked by $\boldsymbol{q}_1, \boldsymbol{q}_2$, and $\boldsymbol{q}_3$ along the three directions related by rotation as indicated by $A - H_\mathrm{I}$, $A - H_\mathrm{II}$, and $A - H_\mathrm{III}$ (Fig. 1g). At first glance, the QPI patterns appear strongly nematic. For example, the separation between $\boldsymbol{q}_1$ and $\boldsymbol{q}_2$ along $A - H_\mathrm{III}$ direction is much larger than that along $A - H_\mathrm{I}$ and $A - H_\mathrm{II}$. This is reminiscent of the pronounced $C_2$-symmetric QPI in the data taken on undoped CsV$_3$Sb$_5$ samples in the CDW phase[13, 14, 22, 29]. However, thanks to our high $q$-space resolution, subtle but distinct QPI differences among all three directions can be resolved, demonstrating that the electronic state in the parent phase indeed breaks all the mirror symmetries in addition to rotational symmetry.

Specifically, the two vectors $\boldsymbol{q}_1$ and $\boldsymbol{q}_2$ overlap along $A - H_\mathrm{I}$ direction, but begin to separate along $A - H_\mathrm{II}$, and clearly form a gap region along $A - H_\mathrm{III}$ (Fig. 1g). These distinctions are much more evident in the 3D zoom-in FT plot (Fig. 2a). Tracking the separation between $\boldsymbol{q}_1$ and $\boldsymbol{q}_2$, we observe increasing distances from $A - H_\mathrm{I}$ to $A - H_\mathrm{III}$ directions. Additional symmetry breaking features are also apparent. The inter-pocket spacing between the triangle-shaped QPI patterns reveals that the distance between $A - H_\mathrm{I}$ and $A - H_\mathrm{II}$ directions ($d_{12}$) is shorter than that between $A - H_\mathrm{II}$ and $A - H_\mathrm{III}$ ($d_{23}$), with the intermediate spacing between $A - H_\mathrm{I}$ and $A - H_\mathrm{III}$ ($d_{13}$). Indeed, the overall QPI map in Fig. 2a clearly breaks all mirror reflection symmetries about the mirror planes $\sigma_v$ and $\sigma'_v$ (Fig. 1a) in terms of both momentum locations in the Brillouin zone and the spectral intensity, attesting to a handedness or chirality of the QPI map and the underlying FS and low energy excitations. Taking the direction of decreasing separation between $q_1$ and $q_2$, which is dominated by the magnitude of $q_2$, as an indicator, the chirality is shown by the red circular arrow (Fig. 2a).

The QPI patterns in the three directions can be directly compared in the side by side view (Fig. 2b) that aligns the Bragg peaks after 60° and 120° rotations. We can clearly identify the misalignment of $q_2$ guided by the grey dashed lines. Moreover, the triangle-shaped QPI features also appear misaligned and distorted, with an evident tilt of the $q_3$ segment. The inner triangular pattern along the $A - H_{\mathrm{II}}$ direction (Fig. 2b, middle) clearly tilts away from the center line, while the outer triangular patterns exhibit an opposite angling. More quantitative visualization of the difference can be revealed by the line profiles (Fig. 2c) along cut#1 (Fig. 2b) in the three $A - H_m$ directions where $m = \mathrm{I, II, III}$. Each line profile exhibits three spectral peaks corresponding to the locations of $q_1$, $q_2$, and $q_3$, which clearly change and confirm the directional variation in the magnitude of $q_1$, $q_2$, and $q_3$. Similarly, the line profiles along $L - H$ in the three $A - H_m$ directions (cut#2 in Fig. 2b) confirm the different spacings between triangle-shaped QPI features (Fig. 2d). We note that chiral QPI patterns with the opposite chirality are observed in another region several hundred nanometers away (Fig. 2g and Extended Data Fig. 1), suggesting the existence of different chiral domains.

These directional and geometric chiral distortions of the QPI patterns suggest underlying chiral distortion of the FSs. To confirm this, we start with the $C_6$-symmetric FSs of both the $\alpha$ and $\delta$ pockets predicted by the DFT (Fig. 1c), extract quantitatively the Fermi wave vectors $k^0_{F,n}(\theta)$ for both the $\alpha$ ($n = 1$) and $\delta$ ($n = 2$) pockets, and replot the two sets of FSs as a function of $\theta$ measured from the crystallographic directions (Fig. 2e, left panel). We then conjecture that the correlation effects beyond the DFT theory can be described by a $d$-wave Pomeranchuk distortion, leading to the prediction of the chiral FSs $k_{F,n}(\theta) = k^0_{F,n}(\theta) + \eta \cdot \cos(2\theta - \pi/6)$, which break all mirror symmetries (Fig. 2e, right panel). The simulated QPI patterns calculated from the joint density of states of both the symmetric and chiral FSs (Fig. 2f) demonstrate how such symmetry breaking impacts the QPI pattern. Remarkably, the calculated QPI patterns from the chiral FSs (Fig. 2f, right panel) well describe the experimental observations, capturing all the characteristics of the novel chiral QPI including the chiral variations under rotation of the separation between the circular and the triangular patterns, the spacing between the triangles, and the shape of the triangles along different directions. This establishes the chiral FSs in the parent state of the kagome metals in the absence of the CDW order. Note that the V out-of-plane $d_{xz/yz}$ orbitals have already lost coherence in the parent state, similar to the CDW states in undoped $CsV_3Sb_5$[22].

In addition, the magnitude of the average chiral distortion near the Fermi level can be estimated by calculating the mean difference between $q_1$, $q_2$, and $q_3$ along the three $A - H_m$ directions (Supplementary Table 1). This value is found to be 1.4%~2.2% of $|Q_{Brag}|$, which is smaller than the recently reported optically- and magnetically-induced distortions of approximately $3\%|Q_{Brag}|$ in $RbV_3Sb_5$[30].

**Chiral-nematic quasiparticle dispersions and mass renormalizations**

To further investigate the underlying chiral Fermi liquid state, we performed energy-dependent measurements of the QPI patterns in order to determine the quasiparticle dispersion and the mass renormalizations near the Fermi level, which are the central properties of the Fermi liquid state. These properties are directly determined by the electron interactions through the Landau Fermi liquid parameters. The line profiles in the Fourier transforms of the d$I$/d$V$ maps are shown in Fig. 3a along cut #1 (Fig. 2b), corresponding to orientations along $A - H_I$ (left), $A - H_{II}$ (middle), and $A - H_{III}$ (right) directions. The three prominent QPI patterns associated with the three scattering wavevectors $q_{1,2,3}$ (Fig. 2a-b) exhibit clear dispersions as a function of the bias-energy (Fig. 3a), reflecting those from scattering of the quasiparticle bands associated with the $\alpha$ and $\delta$ Fermi surfaces (Fig. 3b, inset). The quantitative values of $q_i$ ($i = 1, 2, 3$) have been extracted at each bias-voltage ($E$) by Lorentzian fits (Fig. S1) and superimposed in Fig. 3a as open circles, clearly revealing the dispersion of the scattering wave vectors $q_i(E)$, along each of the three $A - H_m$ directions.

If the electrons were in the symmetric Fermi liquid phase, the quasiparticle energy-momentum dispersion would be described by the DFT electronic structures (Fig. 3b). The $q_i$ ($i = 1, 2, 3$) in the QPI corresponds to the scattering wavevectors connecting the equal-energy contours (insets in Fig. 3b). Comparing to the observed dispersions (Fig. 3a), overall agreement near the Fermi level ($E = 0$) can be seen, including the location and the slope of the dispersion for $q_i(E)$. However, in the symmetric phase predicted by the DFT, the mirror symmetries (Fig. 2c) require $q_i(E)$ to be the same along the three $A - H_m$ directions, which does not agree with the observations (Fig. 3a). The observed dispersions $q_i^m(E)$ in the $A - H_m$ directions are noticeably different and misalign under mirror reflections. This indicates that the quasiparticle dispersions break all mirror symmetry in addition to rotation symmetry, i.e. exhibiting chiral-nematic Fermi liquid behavior.

To investigate more quantitatively the dispersions of the scattering wavevectors, the zoom-in low-energy $q_1^m(E)$ associated with the central $\alpha$ pocket are plotted together in Fig. 3c in the three directions. Approximate linear dispersions $E_s(q_1^m) \simeq v_{1,s}^m(q_1^m - q_{1,F}^m)$ are observed, which originate from those of the Fermi liquid quasiparticles near the Fermi level. The deviations from the linear behavior sufficiently above and below the Fermi level are due to the electron-phonon coupling that produces the "kink" feature observed in the quasiparticle dispersion[29, 31]. Remarkably, the velocities of the scattering dispersion $v_{1,s}^m$ are different in all three $A - H_m$ directions, thus breaking all mirror reflection symmetries with a handedness or chirality $v_{1,s}^{III} < v_{1,s}^{I} < v_{1,s}^{II}$ (Supplementary Table 2). This is in sharp contrast to the prediction of the DFT theory that gives $v_{1,s}^{DFT} = 3037.36$ meV/$|Q_{Brag}|$ in all three directions in the symmetric phase. Because the scattering across the central $\alpha$ band is connected by inversion, which remains a symmetry in the chiral-nematic state, the quasiparticle dispersion along the three directions $E(k_1^m) \simeq v_{1,F}^m(k_1^m - k_{1,F}^m)$ are readily obtained with the Fermi velocities $v_{1,F}^m = 2v_{1,s}^m = (4928.84, 5115.95, 4350.17)$ meV/$|Q_{Brag}|$. Comparing to the DFT-calculated Fermi velocity $v_{1,F}^{DFT} = 6074.72$ meV/$|Q_{Brag}|$, we obtain the chiral mass renormalization ratio $m_1^*/m_1^{DFT} = (1.23, 1.19, 1.40)$ in the three directions for the Sb $p$-orbital derived $\alpha$ band near the Fermi level.

Importantly, the chiral quasiparticles are also observed through $q_{2,(3)}^m(E)$ in the $A - H_m$ directions (Fig. 3a), which are associated with the quasiparticle scattering between the outer Fermi pockets of the $\delta$ band (Fig. 3b) from the V $d$-orbitals. The corresponding energy dispersions (Fig. 3d) consistently show the linear behavior $E_s(q_{2(3)}^m) \simeq v_{2(3),s}^m(q_{2(3)}^m - q_{2(3),F}^m)$ near the Fermi level with deviations above and below due to the electron-phonon coupling that changes the quasiparticle dispersions involved in the elastic scattering process. Once again, the scattering velocities $v_{2(3),s}^m$ are different in the three $A - H_m$ directions, breaking all mirror reflection symmetries. The magnitude of the velocities exhibit handedness or chirality $v_{2,s}^{I} > v_{2,s}^{II} > v_{2,s}^{III}$ and $v_{3,s}^{I} > v_{3,s}^{II} > v_{3,s}^{III}$ due to electron correlations (Supplementary Table 2), in contrast to the DFT results where $v_{2,s}^{DFT} = 1263.32$ and $v_{3,s}^{DFT} = 1607.08$ meV/$|Q_{Brag}|$ in all three directions. Different from the central $\alpha$ band, the scattering states connected by $q_{2,3}^m$ are not related by inversion and are affected by the chiral-nematic distortion (Fig. 2c). As a result, the quasiparticle dispersions $E(k_{2(3)}^m) \simeq v_{2(3),F}^m(k_{2(3)}^m - k_{2(3),F}^m)$ cannot be accurately determined from that of the quasiparticle scattering wavevector. However, to leading order in the chiral-nematic distortion, $v_{2(3),F}^m \simeq 2v_{2(3),s}^m$ still holds, which allows a good estimate of the renormalized Fermi velocities: $v_{2,F}^m =$

(1554.41, 1279.89, 1074.19) and $v_{3,F}^m$ = (2571.86, 1979.03, 1703.77) meV/$|Q_{Brag}|$, and the mass renormalizations $m_{2,m}^*/m_2^{DFT}$ = (1.63, 1.97, 2.35) and $m_{3,m}^*/m_3^{DFT}$ = (1.25, 1.62, 1.89) along the three directions for the δ band associated with the V $d$-orbitals.

The extracted mass renormalization ratios for most bands fall within the range of 1.29–1.71 predicted by DFT+DMFT[32, 33]. However, the δ band exhibits slightly higher renormalization along specific directions, reinforcing the existence of an exotic state that breaks mirror and rotational symmetries. Furthermore, the reported ARPES data[31] show a similarly enhanced renormalization for the δ band compared to the theoretical prediction, in agreement with our QPI results.

**Chiral-nematic residual Fermi surface in the presence of CDWs**

We have shown that CsV$_{3-x}$Ti$_x$Sb$_5$ at $x$ = 0.15 is a novel chiral-nematic Fermi liquid, where both the FSs and the quasiparticle excitations break rotation and mirror symmetries and exhibit handedness or chirality. The symmetry breaking is surprisingly strong and suggests that the chiral-nematic Fermi liquid may emerge at temperatures above the CDW transitions and thus impact the physical properties of the CDW state observed at small $x$ including the stochiometric AV$_3$Sb$_5$. To explore the lightly doped CsV$_{3-x}$Ti$_x$Sb$_5$, we focus on the $x$ = 0.05 sample, in which the CDWs are only partially suppressed (Fig. 1**b**, gray arrow). The d$I$/d$V$($r$, -5 mV) map reveals apparent unidirectional stripes (Fig. 4**a**). The corresponding FT d$I$/d$V$($q$, -5 mV) exhibits additional features compared to the $x$ = 0.15 sample (Fig. 2**a**), including peaks associated with the $2a_0 \times 2a_0$ CDW and the $4a_0 \times 1a_0$ charge modulations (Fig. 4**b**). Interestingly, these features of the CDW order coexist with two kinds of QPI patterns. First, there are strong unidirectional QPI patterns[13] enclosed by the gray dashed boxes (Fig. 4**b**), which most likely originate from the reconstructed FSs due to charge stripe order. Surprisingly, there are also QPI patterns associated with the residual FSs in the CDW state indicated by the $q_1$, $q_2$, and $q_3$ segments (Fig. 4**b**), that closely resemble the corresponding patterns observed in the $x$ = 0.15 sample without CDW order. As revealed by the 3D zoom-in FT plot (Fig. 4**c**), these QPI patterns display clear directional dependence: the scattering wavevectors $q_1$ and $q_2$ coalesce along the $A - H_I$ direction, gradually separate along the $A - H_{III}$ direction, and clearly split along the $A - H_{II}$ direction. Rotation and mirror symmetries are broken and the QPI patterns are chiral with the indicated handedness.

The QPI patterns can be directly compared in the side by side view (Fig. 4d) that aligns the Bragg peaks in the three directions after 60° and 120° rotations, similar to the analysis performed at $x = 0.15$ (Fig. 2b). The misalignment of $q_2$ guided by the grey dashed line is clearly visible and the triangular QPI features appear distorted from symmetric triangles with a noticeable tilt of the $q_3$ segment. Line profiles (cut#1) extracted from $dI/dV(q, -5\ mV)$ further confirm the directional variation in the lengths of $q_1$, $q_2$, and $q_3$ (Fig. 4e), which closely resembles the behaviors observed in the CDW-free sample at $x=0.15$. These observations of rotation and mirror symmetry breaking suggest that the CDW order may have emerged from the chiral-nematic Fermi liquid parent state, providing novel physical insights for the origin of the chiral CDW order[20, 24, 30, 34]. Indeed, the intensity of the CDW peaks at the wavevectors of the $2a_0 \times 2a_0$ CDW order (Fig. 4b-c) follows a ratio of 2.6:1.2:1.0 and breaks all mirror symmetry, which is indicative of the handedness of the CDW order. We stress that our unprecedented observation of the chiral-nematic Fermi liquid in the absence of the CDW and the chiral CDW state are based on direct measurement of the chiral-nematic electronic structure, i.e. the rotation and mirror symmetry breaking of the Fermi surfaces and the quasiparticle dispersions.

**Discussions**

We have consistently observed the chirality of the electronic structure in $CsV_{3-x}Ti_xSb_5$ through the reproducible chiral QPI patterns at different energies and across various Ti doping levels: low concentration at $x = 0.05$ (Fig. 4), intermediate at $x = 0.15$ (Fig. 2, Figs. S2 and S3), and high concentration at $x = 0.27$ (Extended Data Fig. 2). This suggests that the rotation and mirror symmetry breaking is present independent of the Ti doping level. Although the CDW order gaps out a significant portion of the low-energy V $d$-electron states, making the associated QPI features near the FS weaker, we can still observe the signatures of the chiral FS in the pristine $CsV_3Sb_5$. The QPI measurements on two $x = 0$ samples (Extended Data Figs. 3 and 4) show that the scattering vector $q_1$, associated with Sb $p_z$ orbital, remains clearly visible, while the weaker QPI associated with $q_3$, from the V in-plane $d$ orbital, is faint but still detectable. Both the length variations in $q_1$ along the three different directions (Extended Data Figs. 3e and 4e), as well as the tilt of the $q_3$ feature (Extended Data Figs. 3d and 4d), can be observed, which are consistent with the behavior in the Ti doped samples.

These results support that the chiral-nematicity is a robust and universal feature of the quasiparticles in the $CsV_3Sb_5$ kagome superconductors at all Ti doping, independent of the CDW transition. It is thus reasonable to propose the existence of a chiral-nematic Fermi liquid phase that emerges at a characteristic temperature $T^*$ as the parent state above the CDW transition $T_{CDW} < T^*$ in the AV$_3$Sb$_5$ compounds. From the extracted chiral-nematic distortion of the quasiparticle dispersions of the $\delta$ pockets corresponding to the V $d$-orbitals at $x$=0.15 sample, we estimate an average energy splitting of ~15.0 meV (Supplementary Table 1), corresponding to $T^* \sim 174$ K, which is significantly higher than the highest CDW transition temperature $T_{CDW} \sim 90$ K in the undoped CsV$_3$Sb$_5$. Recently, thermodynamic torque measurements[35] have reported evidence for spontaneous rotation symmetry breaking above the bulk CDW transition and below a characteristic temperature around 130 K. Our observation of the rotation and mirror symmetry breaking electronic structure, independent of the incipience of the CDW order and having an associated energy scale much larger than the thermal energy scale at $T_{CDW}$, provide microscopic evidence for a novel correlated electron state – a chiral-nematic Fermi liquid – already present above the CDW transition.

We thus propose the schematic phase diagram for the kagome metals (Fig. 4f), where the chiral-nematic Fermi liquid emerges from the high-temperature symmetry phase at the temperature $T^*$ due to electron correlations. Below $T^*$, the cascade of exotic CDW and SC phases could originate from the parent chiral-nematic Fermi liquid, inheriting the chiral and nematic characters, consistent with the observation of the chiral CDW phases of the lightly Ti doped sample at $x = 0.05$ (Fig. 4a-e) and the two undoped samples at $x = 0$ (Extended Data Figs. 3 and 4). Moreover, the chirality defined by the all mirror breaking pair density modulations has been observed recently in the SC state of (Cs,K)V$_3$Sb$_5$ by Josephson tunneling STM[18]. We stress again that "chirality" here refers to breaking all mirror symmetries and does not imply the time-reversal symmetry breaking which further depends on the tunability of chirality by applied magnetic field[18, 20, 30].

To further test this physical idea, we performed Bogoliubov QPI (B-QPI) measurements in the $x = 0.15$ sample (Extended Data Fig. 5), where the CDW order is completely suppressed and the parent chiral-nematic Fermi liquid transitions directly into the SC state (Fig. 4f) with a U-shaped SC gap in the d$I$/d$V$ spectra (Fig. S4 and Extended Data Fig. 5d). By measuring the d$I$/d$V$ map at the subgap energy (Extended Data Fig. 5b), where the QPI is primarily governed by the Bogoliubov quasiparticles, we find that the overall B-QPI patterns closely resemble those of the normal state (Extended Data Fig. 5c)[22]. These include the misalignment of $q_2$ in the $A - H_I$, $A - H_{II}$, and $A - H_{III}$ directions and the tilt of $q_3$ (Extended Data Fig. 5e), different $q_1$, $q_2$, and $q_3$ lengths along the three directions in the extracted line profiles (Extended Data Fig. 5f), as well as the similar chiral features identified from the 3D zoom-in FT plot (Extended Data

Fig. 5g) compared to the normal state QPI (Fig. 2a). Furthermore, the chiral nature of the QPI patterns are also observed across the superconducting transition temperature $T_c$ (3.5 K[28]), both below (0.4 K) and above (4 K) (Figs. S2 and S3). The broken mirror and rotation symmetries of the B-QPI are strong evidence for a chiral-nematic SC ground state[36-38] in $CsV_{3-x}Ti_xSb_5$ at $x = 0.15$. This finding is new and attests to the existence of the parent chiral-nematic Fermi liquid as the origin of robust chiral low-temperature states observed in $CsV_3Sb_5$.

Finally, we propose a plausible theoretical picture for the emergence of the rotation and mirror symmetry breaking chiral-nematic Fermi liquid. In the presence of translation symmetry, the rotation symmetry breaking can arise from a Pomeranchuk instability of the symmetric Fermi liquid due to quadrupolar order. Such a $d$-wave Pomeranchuk instability has been widely studied in systems with isotropic or fourfold-symmetric Fermi surfaces, where it typically leads to nematic FS distortions[11,39-42]. On the kagome lattice, a general FS with $C_6$ rotational symmetry ($C_3 \oplus$ inversion) can be described by $k_F^0(\theta) = a \cdot \cos(6\theta) + b$ inside a hexagonal Brillouin zone (Extended Data Fig. 6a) where $a$ and $b$ are constants and the angle $\theta$ is measured from the crystalline directions. Two sets of mirror plane perpendicular to the kagome plane are indicated by $\sigma_v$ and $\sigma_v'$, corresponding to $\theta = n\pi/6, n = 0, 1, ... 11$. A conventional $l = 2$ Pomeranchuk distortion described by $\eta \cdot \cos 2\theta$ can only produce a nematic FS since its two mirror arises coincide with $\sigma_v$ and $\sigma_v'$ and thus does not break all mirror symmetries. To describe the chiral-nematic distortion, the quadrupolar order must be phase-shifted from the crystalline directions, such that the distorted FS $k_F(\theta) = k_F^0(\theta) + \eta \cdot \cos(2\theta + \varphi)$ under a generic $\varphi$ can produce the chiral-nematic FS that breaks rotation and all mirror reflections $\sigma_v$ and $\sigma_v'$ (Extended Data Fig. 6a). Indeed, it is both intriguing and exciting that the observed QPI patterns (Fig. 2) can be very well captured elegantly by this simple expression with $k_F^0(\theta)$ corresponding to the DFT results for the FSs and the quadrupolar phase twist $\varphi = -\pi/6$ used for our QPI data analysis (Fig. 2e-f).

The microscopic origin for the chiral-nematic quadrupolar order demands theoretical studies. In principle, both lattice and electronic channels could generate such a state. We conjecture that, in the electronic channel, an intra-unit-cell, sublattice charge order can naturally explain its origin (see Method and Extended Data Fig. 6b). Specifically, the charge densities on the three V atoms on the three sublattices are driven by correlations to be different, which may also be accompanied by three different bonds connecting the sublattices. Below the temperature $T^*$, the kagome metal transitions from the symmetric FL to the chiral-nematic FL with sublattice charge order (Fig. 4f), impacting the subsequent translation symmetry breaking associated with the CDW order and/or the U(1) symmetry breaking associated with the SC order.

**Summary**

By direct visualization of the electronic states, we discovered a novel chiral-nematic Fermi liquid state in kagome metals $CsV_{3-x}Ti_xSb_5$ in the absence of CDW order (large $x$), with clear manifestations in the chiral CDW metal phase (small $x$) and/or the SC states (all $x$), marking significant advances in understanding the origin of the exotic chirality/handedness in kagome materials. The onset temperature $T^*$ of this parent state may be precisely determined in the future by temperature-dependent microscopic probes, such as nano-ARPES in resolved chiral domains. Our results not only provide a plausible chiral-nematic *parent state* due to correlated quadrupolar electronic order, but also establish a new conceptual framework for exploring the emergence and consequences of such exotic quantum phases due to mirror and rotation symmetry breaking. Our findings offer critical insights into the fundamental origin of chirality in kagome metals and superconductors and other two-dimensional quantum materials.

## Materials and Methods

**Single crystal growth of the Ti-doped CsV$_3$Sb$_5$ samples.** Single crystals of Ti-doped CsV$_3$Sb$_5$ were grown from Cs liquid (purity 99.98%), V powder (purity 99.9%), Ti shot (purity >99.9%) and Sb shot (purity 99.999%) *via* a modified self-flux method[14,28]. Single crystals of CsV$_3$Sb$_5$ were grown from Cs liquid (purity 99.98%), V powder (purity 99.9%) and Sb shot (purity 99.999%) *via* a modified self-flux method[14,28].

**Scanning tunneling microscopy/spectroscopy.** The samples used in the STM/S experiments were cleaved at low temperature (13 K) and immediately transferred to an STM chamber. Experiments were performed in an ultrahigh vacuum (1×10$^{-10}$ mbar) ultra-low temperature STM system equipped with 11 T magnetic field. All the scanning parameters (setpoint voltage and current) of the STM topographic images are listed in the figure captions. The base temperature is 0.4 K in the low-temperature STS and the electronic temperature is 650 mK, calibrated using a standard superconductor, Nb crystal. Unless otherwise noted, the d$I$/d$V$ spectra were acquired by a standard lock-in amplifier at a modulation frequency of 973.1 Hz, and all the data in main text were measured at 0.4 K. Non-superconducting tungsten tips were fabricated via electrochemical etching and calibrated on a clean Au(111) surface prepared by repeated cycles of sputtering with argon ions and annealing at 500 °C. The Cs adatoms at as-cleaved Sb surface were moved away by the STM tip to form a large-scale and clean Sb surface[14]. In Figs. 2**b** and 4**d**, to eliminate the drift effects inherent in STM scanning, an affine transformation algorithm is applied to the Bragg points to restore a perfectly symmetric lattice[43]. To improve the signal-to-noise ratio, the extracted line profiles are averaged by summing the six neighboring lines along the corresponding symmetry direction.

**Theoretical calculations.** Calculations were performed within the density-functional theory (DFT) as implemented in VASP package[44]. The generalized-gradient-approximation as parametrized by Perdew-Burke-Ernzerhof[45] for the exchange-correlation interaction between electrons was employed in all calculations. Zero damping DFT-D3 vdW correction[46] was also employed in all calculations while spin-orbital coupling was not included. A cutoff energy of 300 eV for the plane-wave basis set was used. Bulk band structures of Ti doped CsV$_3$Sb$_5$ are calculated with the virtual crystal approximation[47]. The QPI simulation is then carried out using an auto-correlation algorithm[13], based on those defined Fermi surfaces.

**Possible origin of chiral-nematic charge order.** The symmetric lattice and chiral-nematic charge order are shown in Extended Data Fig. 6b, respectively. The onsite charge density around a hexagon is expressed as a function of the angular coordinate $\theta$, measured from the $x$-direction,

$$n(\theta) = n_0 \cos 6\theta + \delta n \cos(2\theta + \varphi),$$

where $\theta = 0, \pm\frac{\pi}{3}, \pm\frac{2\pi}{3}, \pi$ correspond to the six sites and $\{n_0, \delta n, \varphi\}$ parameterize the charge distribution. Because inversion symmetry is preserved, the six sites fall into three pairs as labeled in Extended Data Fig. 6b, with charge densities

$$n_1 = n_0 + \delta n \cos \varphi,$$

$$n_2 = n_0 - \frac{1}{2}\delta n \cos \varphi - \frac{\sqrt{3}}{2}\delta n \sin \varphi,$$

$$n_3 = n_0 - \frac{1}{2}\delta n \cos \varphi + \frac{\sqrt{3}}{2}\delta n \sin \varphi.$$

When $\delta n = 0$, the charge distribution $n_i = n_0$ is six-fold symmetric, as shown in the left panel of Extended Data Fig. 6b, corresponding to the state above $T^*$. When $\delta n \neq 0$ and $\varphi = 0, \pm\pi$, one obtains a nematic charge order $n_1 \neq n_2 = n_3$, retaining two mirror planes along the $x$- and $y$-direction of the lattice. For $\varphi = \pm\pi/2$, the nematic component vanishes and the charge order becomes purely chiral, with $n_1 - n_2 = n_3 - n_1$. For generic $\varphi \neq 0, \pm\pi$, all mirror symmetries are broken and the pattern becomes chiral-nematic, characterized by $n_1 \neq n_2 \neq n_3$, as illustrated in the right panel of Extended Data Fig. 6b.

## Author Contributions:

# Main Figures

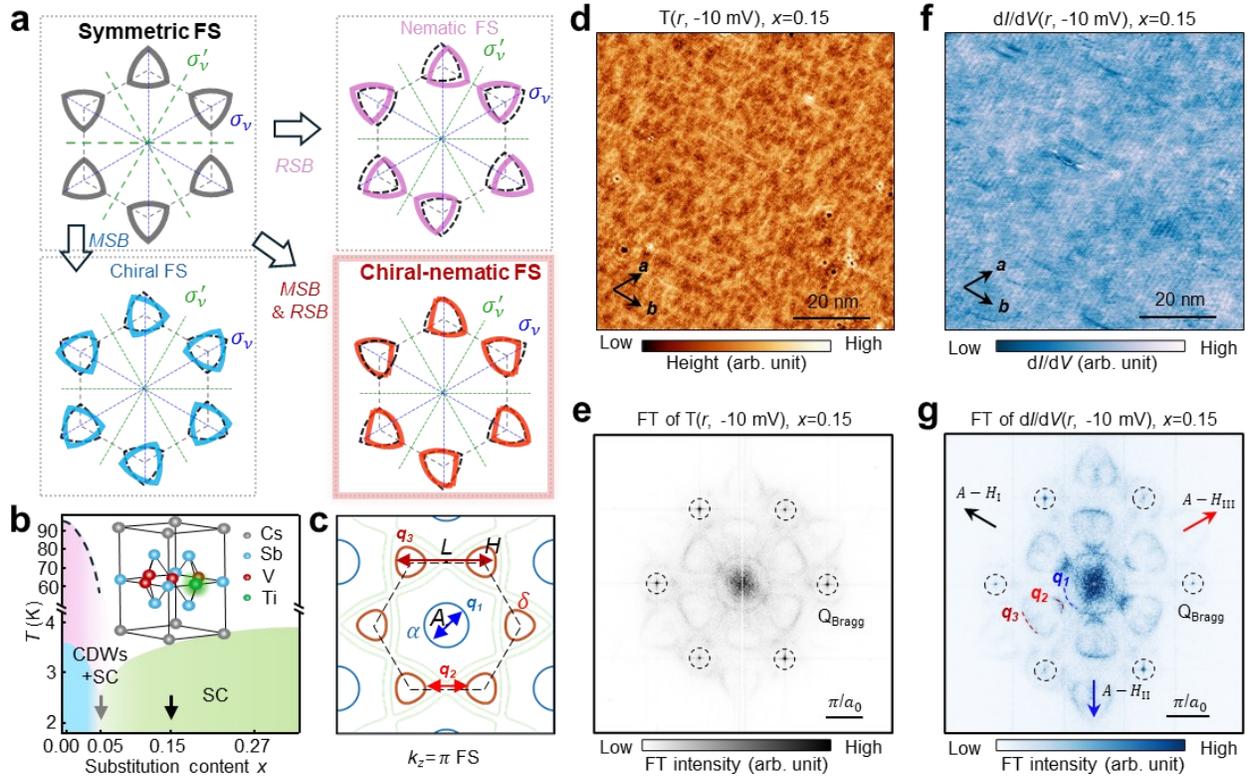

**Fig. 1. Symmetry-breaking Fermi surfaces and orbital-resolved QPI in Ti-doped $CsV_3Sb_5$ without CDW order**. **a**, Schematic of FS renormalization of the chiral, nematic, and chiral-nematic electronic states, respectively. Two sets of mirror planes, labeled $\sigma_v$ and $\sigma_v'$, are indicated. RSB: Rotational symmetry breaking. MSB: Mirror symmetry breaking. **b**, Phase diagram of $CsV_{3-x}Ti_xSb_5$, highlighting the non-CDW region at $x = 0.15$. Inset is the schematic crystal structure of $CsV_{3-x}Ti_xSb_5$. **c**, DFT-calculated CEC ($k_z=\pi$) at Fermi level in $CsV_{3-x}Ti_xSb_5$ ($x=0.15$) showing the multiorbital FSs. Arrows indicate scattering vectors corresponding to $q_1$, $q_2$, and $q_3$, respectively. **d**, STM topographic image of $CsV_{3-x}Ti_xSb_5$ ($x=0.15$). Scanning parameters: $V_{bias}$= -10 mV, $I$= 500 pA. **e**, FT of (**d**), showing the absence of the CDW. **f**, d$I$/d$V$ map of the same area with (**d**). $V_{bias}$= -10 mV, $I$= 500 pA, $V_{lock-in}$= 0.5 mV. **g**, FT of (**f**) showing QPI patterns. Bragg peaks are marked by black dashed circles. Arrows denote the $A-H_I$, $A-H_{II}$, and $A-H_{III}$ directions, and QPI vectors $q_1$, $q_2$, and $q_3$ are indicated by dark red dashed curves, respectively.

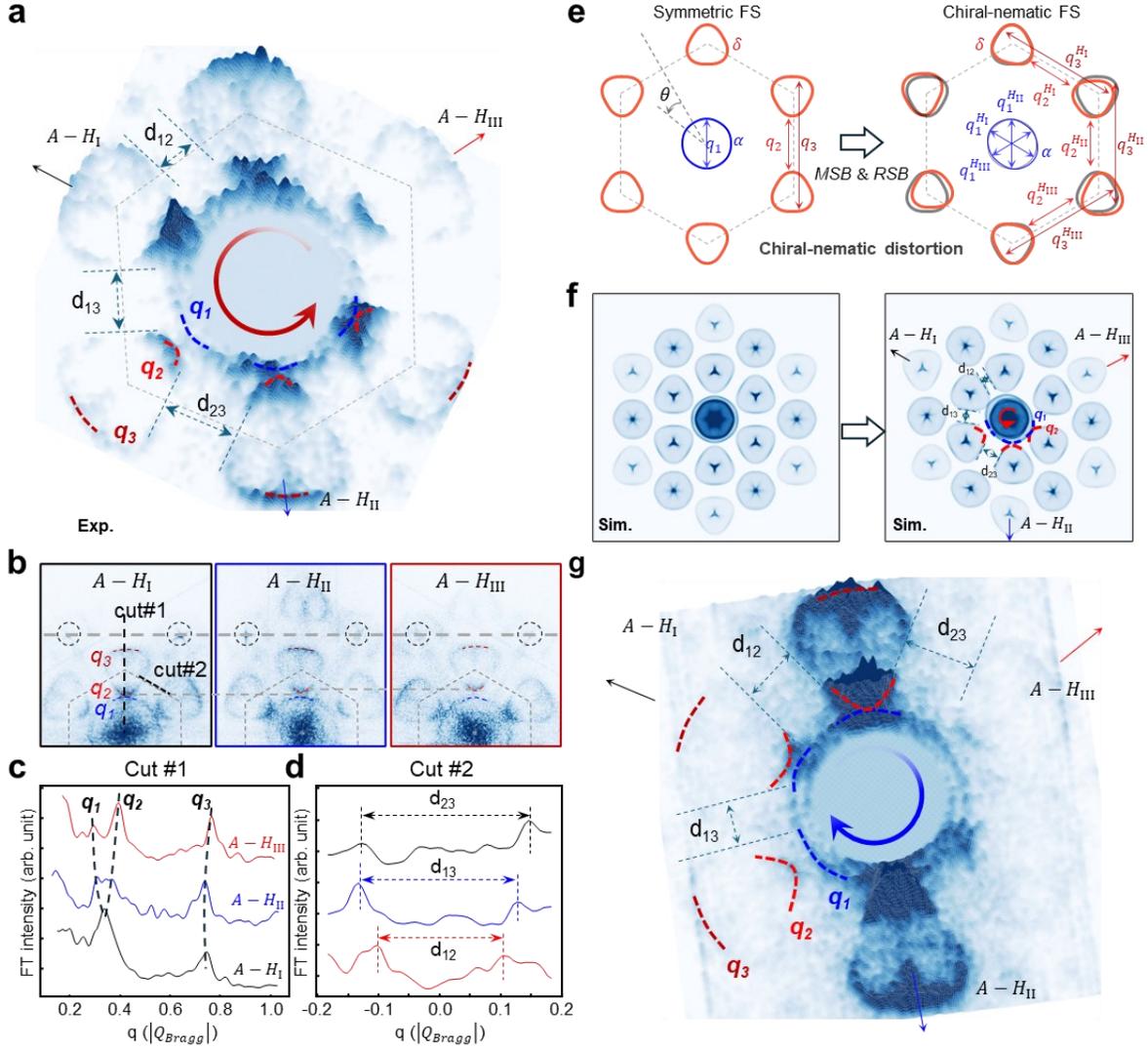

**Fig. 2. Observation of chiral-nematic Fermi surface in Ti-doped CsV$_3$Sb$_5$ without CDW order**. **a**, 3D plot of the zoomed-in FT of the d$I$/d$V$ map in CsV$_{3-x}$Ti$_x$Sb$_5$ ($x$=0.15), highlighting the QPI patterns. The curved arrow in red shows the chirality of the QPI patterns. The central region is masked for clarity. Scanning parameters: $V_{bias}$= -10 mV, $I$= 500 pA, $V_{lock-in}$= 0.5 mV. **b**, QPI patterns along the $A-H_I$, $A-H_{II}$, and $A-H_{III}$ directions, respectively, highlighting the difference in three directions and showing the mirror symmetry breaking of the QPI patterns. **c**, Line profiles (cut#1) along the $A-H_I$, $A-H_{II}$, and $A-H_{III}$ directions in (**a**), respectively, quantitively showing the chirality of $q_i^{H_m}$ ($i$ = 1,2,3; $m$ = I, II, III). The chirality is indicated by dotted curves. **d**, Line profiles (cut#2) along the different $L-H$ directions, showing the chiral distance between the triangular QPI features. **e**, Schematic of the chiral Fermi surface based on $\alpha$ and $\delta$ pockets, showing mirror and rotational symmetry breaking. The scattering vectors $q_1$, $q_2$, and $q_3$ in different directions are labeled as $q_i^{H_m}$ ($i$ = 1,2,3; $m$ = I, II, III), respectively. **f**, Simulated QPI patterns based on symmetric and chiral Fermi surfaces in (**e**), showing distorted QPI pattern, which agrees with the observed distorted QPI results in (**a**). **g**, 3D plot of zoomed-in FT of d$I$/d$V$ map from another region of the same sample, revealing the opposite chirality in the QPI patterns compared to (**a**). $V_{bias}$= -10 mV, $I$= 1000 pA, $V_{lock-in}$= 0.3 mV.

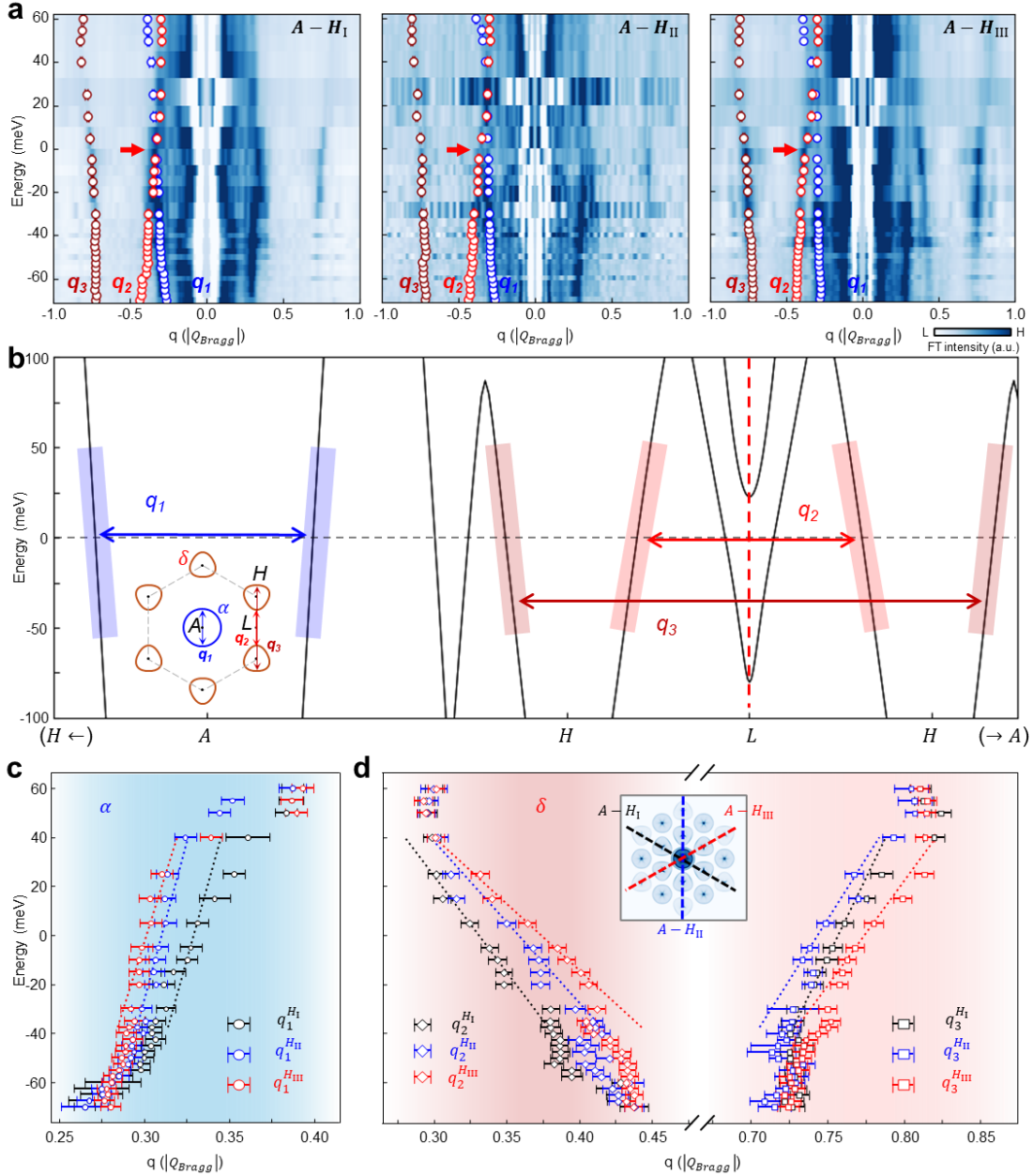

**Fig. 3. Observation of chiral-nematic quasiparticle dispersions and renormalizations in Ti-doped CsV$_3$Sb$_5$ without CDW order**. **a**, Bias-dependent QPI line profiles extracted from the FTs of d$I$/d$V$ maps along the $A - H_\mathrm{I}$ (left), $A - H_\mathrm{II}$ (middle), and $A - H_\mathrm{III}$ (right) directions (cut#1), respectively, showing the dispersions of $q_1$, $q_2$, and $q_3$. The colored open circles highlight the quantitative values of $q_1$, $q_2$, and $q_3$ extracted from QPI line profiles. The red arrows highlight the difference around the Fermi level. Fitted Gauss-like backgrounds are subtracted for better view. **b**, DFT-calculated electronic structure at $k_z=\pi$, showing the bands from which the $q_1$, $q_2$, and $q_3$ originate, as marked by arrowed lines in blue and red. Inset is the corresponding Fermi surface with q-vectors indicated. **c**, **d**, Dispersions $E_s(q_1)$ (**c**), $E_s(q_2)$, and $E_s(q_3)$ (**d**) extracted from (**a**) along $A - H_\mathrm{I}$, $A - H_\mathrm{II}$, and $A - H_\mathrm{III}$ directions, respectively, showing the chiral Fermi velocity renormalizations at low energy. Dotted lines show the linear fittings of the dispersions at low energy.

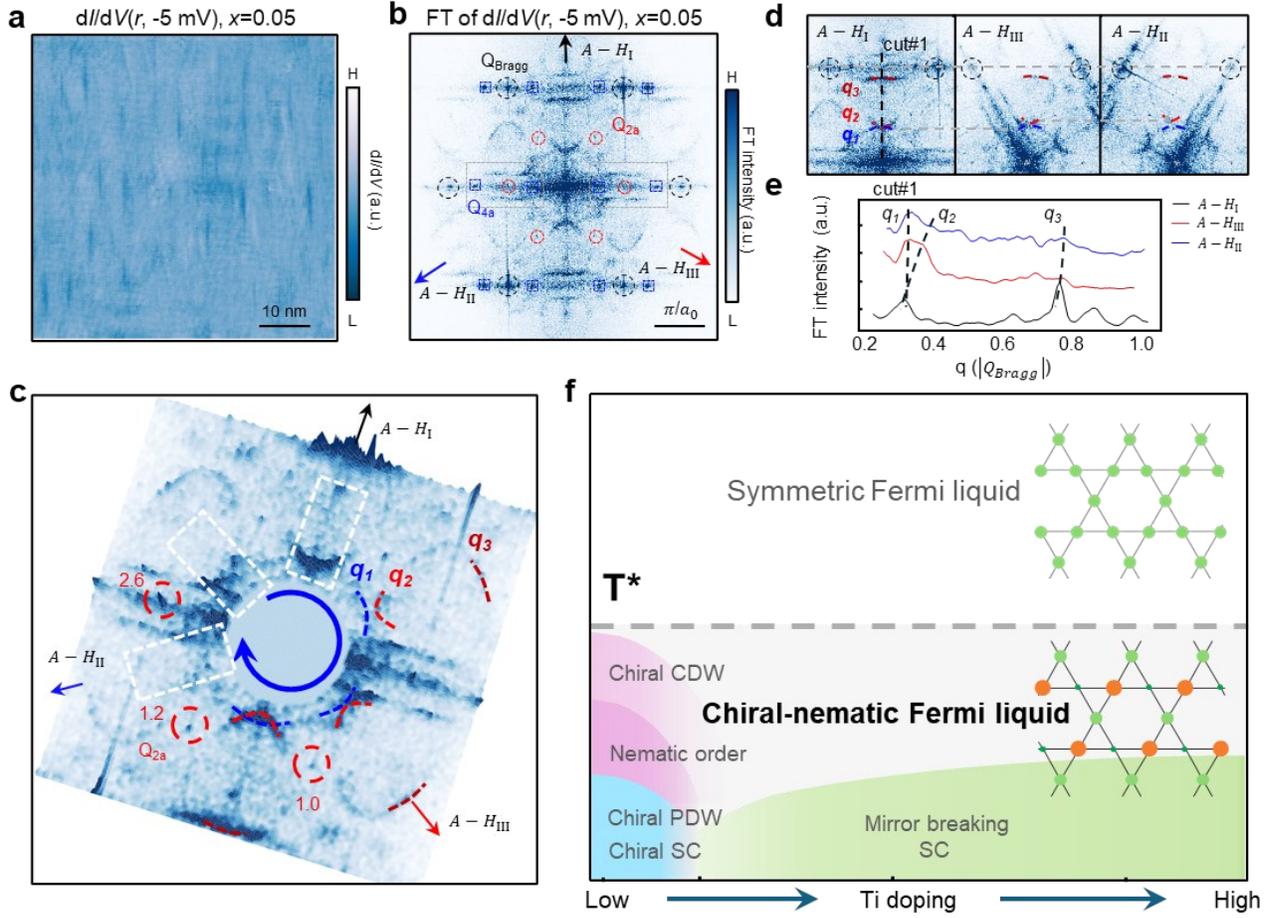

**Fig. 4. Observation of chiral-nematic residual Fermi surface in lightly Ti-doped CsV$_3$Sb$_5$ with CDW order**. **a**, d$I$/d$V$ map of CsV$_{3-x}$Ti$_x$Sb$_5$ ($x$=0.05), showing the presence of charge orders. $V_{bias}$= -5 mV, $I$= 200 pA, $V_{lock-in}$= 0.2 mV. **b**, FT of (**a**), showing the presence of multiple electronic states. Black dotted circles, red dotted circles, and blue dotted rectangles mark the Bragg points, $2a_0 \times 2a_0$ CDW order, and $1a_0 \times 4a_0$ stripes order, respectively. $A - H_{\mathrm{I}}$, $A - H_{\mathrm{II}}$, and $A - H_{\mathrm{III}}$ directions are indicated by arrows. **c**, 3D plot of the zoomed-in FT in (**b**), highlighting the QPI pattern. The curved arrow denotes the chirality of QPI patterns. White dashed rectangles highlight the evolving separation between $q_1$ and $q_2$ vectors. The intensity ratios of the $2a_0 \times 2a_0$ CDW peaks as marked by red dotted circles is labeled. The central region is masked for clarity. **d**, QPI pattern along $A - H_{\mathrm{I}}$, $A - H_{\mathrm{II}}$, and $A - H_{\mathrm{III}}$ directions, respectively, highlighting the difference in three directions and showing the mirror symmetry breaking of the QPI pattern. **e**, Line profiles (cut#1) along $A - H_{\mathrm{I}}$, $A - H_{\mathrm{II}}$, and $A - H_{\mathrm{III}}$ directions in (**b**), respectively, showing the chirality of $q_i^{Hm}$ (i = 1,2,3; m = I, II, III). The different lengths of scattering vectors are indicated by dotted curves. **f**, Schematic phase diagram for kagome metals and superconductors, illustrating the emergence of the chiral-nematic Fermi liquid state as a parent state in CsV$_{3-x}$Ti$_x$Sb$_5$. Inset shows a uniform charge distribution (upper right) corresponding to symmetric Fermi liquid and an intra-cell, sublattice charge order (lower right) corresponding to chiral-nematic Fermi liquid.

# Extended Data Figures

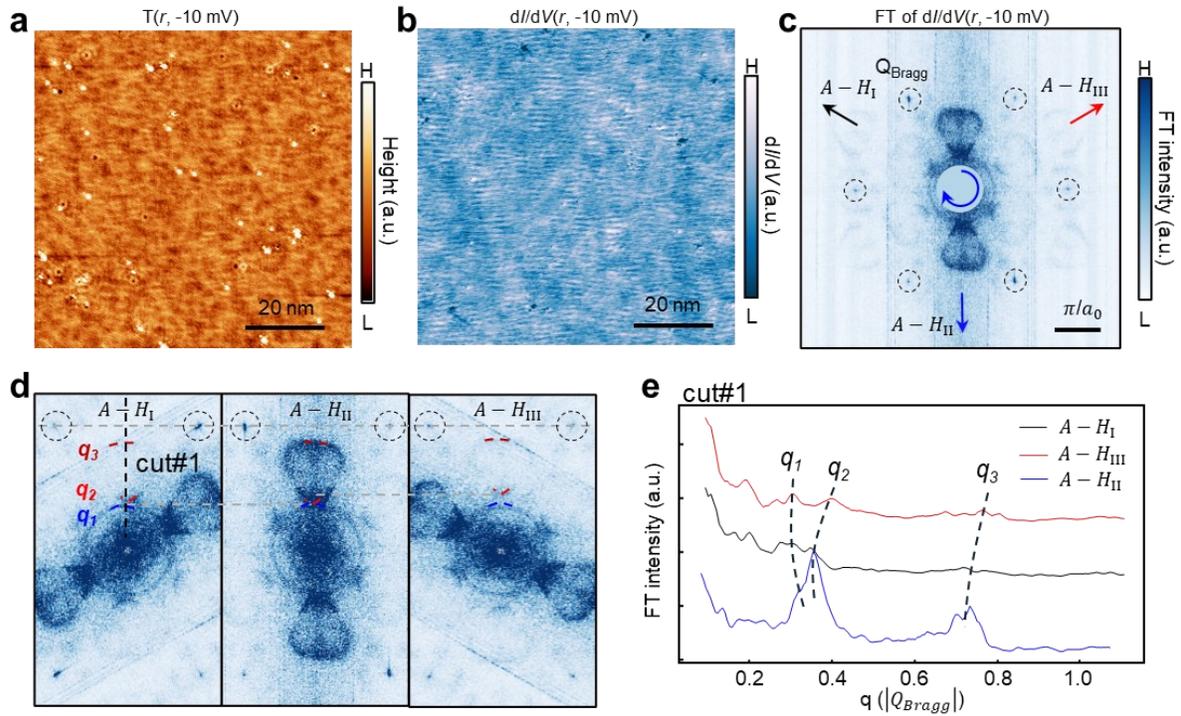

**Extended Data Fig. 1. Opposite chiral QPI observed in CsV$_{3-x}$Ti$_x$Sb$_5$ ($x$=0.15) at a region near the area shown in Fig. 2**. **a**, STM topographic image of CsV$_{3-x}$Ti$_x$Sb$_5$ ($x$=0.15). $V_{bias}$= -10 mV, $I$= 1000 pA. **b**, d$I$/d$V$ map of (**a**). $V_{bias}$= -10 mV, $I$= 1000 pA, $V_{lock-in}$= 0.3 mV. **c**, FT of (**b**). Black dotted circles mark the scattering vectors of Bragg points. $A-H_I$, $A-H_{II}$, and $A-H_{III}$ directions are indicated by black, blue, and red arrows, respectively. The curved arrow denotes the opposite chirality compared to the region of Fig. 2. **d**, The QPI patterns along $A-H_I$, $A-H_{II}$, and $A-H_{III}$ directions, highlighting the difference in three directions and showing the mirror symmetry breaking of the QPI patterns. **e**, Line profiles (cut#1) extracted along $A-H_I$, $A-H_{II}$, and $A-H_{III}$ directions in (**c**), showing the different lengths of scattering vectors $q_1$, $q_2$, and $q_3$, respectively, as indicated by dotted curves.

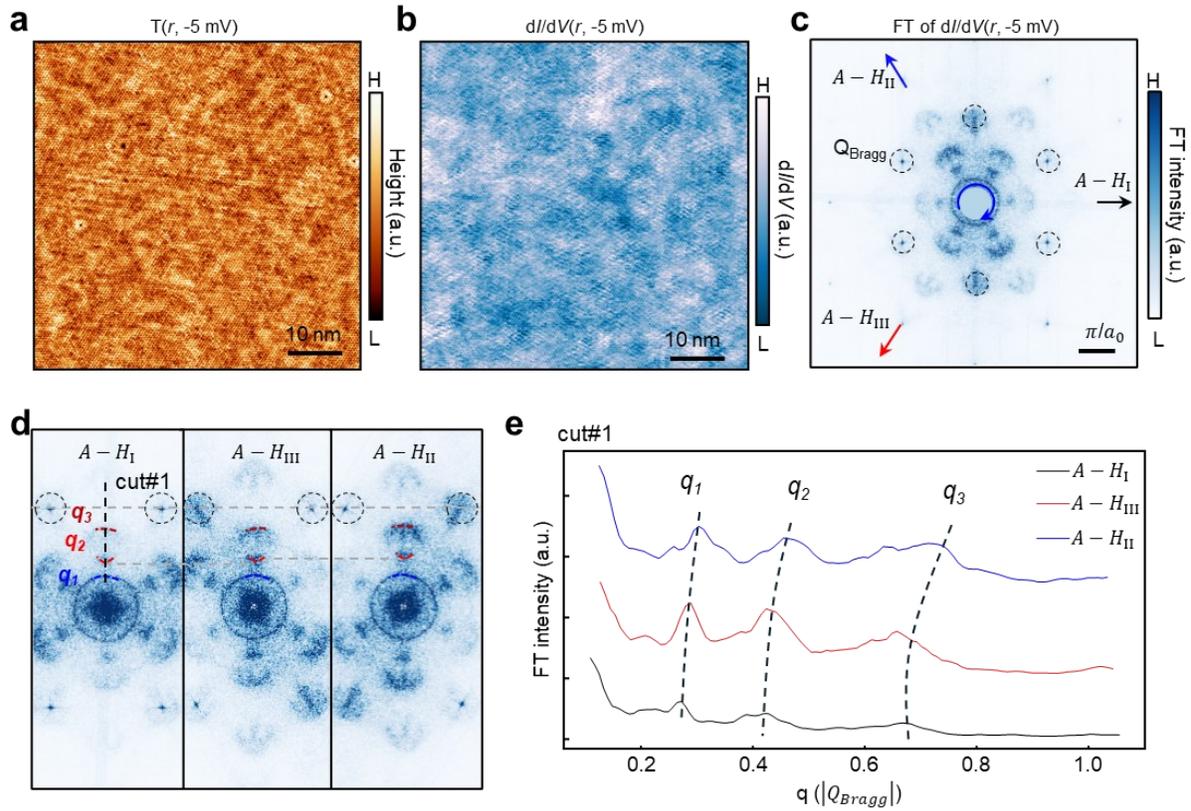

**Extended Data Fig. 2. Chiral QPI in CsV$_{3-x}$Ti$_x$Sb$_5$ (x=0.27) at 0.4K.** **a**, STM topographic image of CsV$_{3-x}$Ti$_x$Sb$_5$ (x=0.27). $V_{bias}$= -5 mV, $I$= 500 pA. **b**, d$I$/d$V$ map of (**a**). $V_{bias}$= -5 mV, $I$= 500 pA, $V_{lock-in}$= 1 mV. **c**, FT of (**b**). Black dotted circles mark the scattering vectors of Bragg points. $A-H_I$, $A-H_{II}$, and $A-H_{III}$ directions are indicated by black, blue, and red arrows, respectively. The curved arrow denotes the chirality. **d**, The QPI patterns along the $A-H_I$, $A-H_{II}$, and $A-H_{III}$ directions, highlighting the difference in three directions and showing the mirror symmetry breaking of the QPI patterns. **e**, Line profiles (cut#1) extracted along $A-H_I$, $A-H_{II}$, and $A-H_{III}$ directions from (**c**), showing the different lengths of scattering vectors $q_1$, $q_2$, and $q_3$, respectively, as indicated by dotted curves.

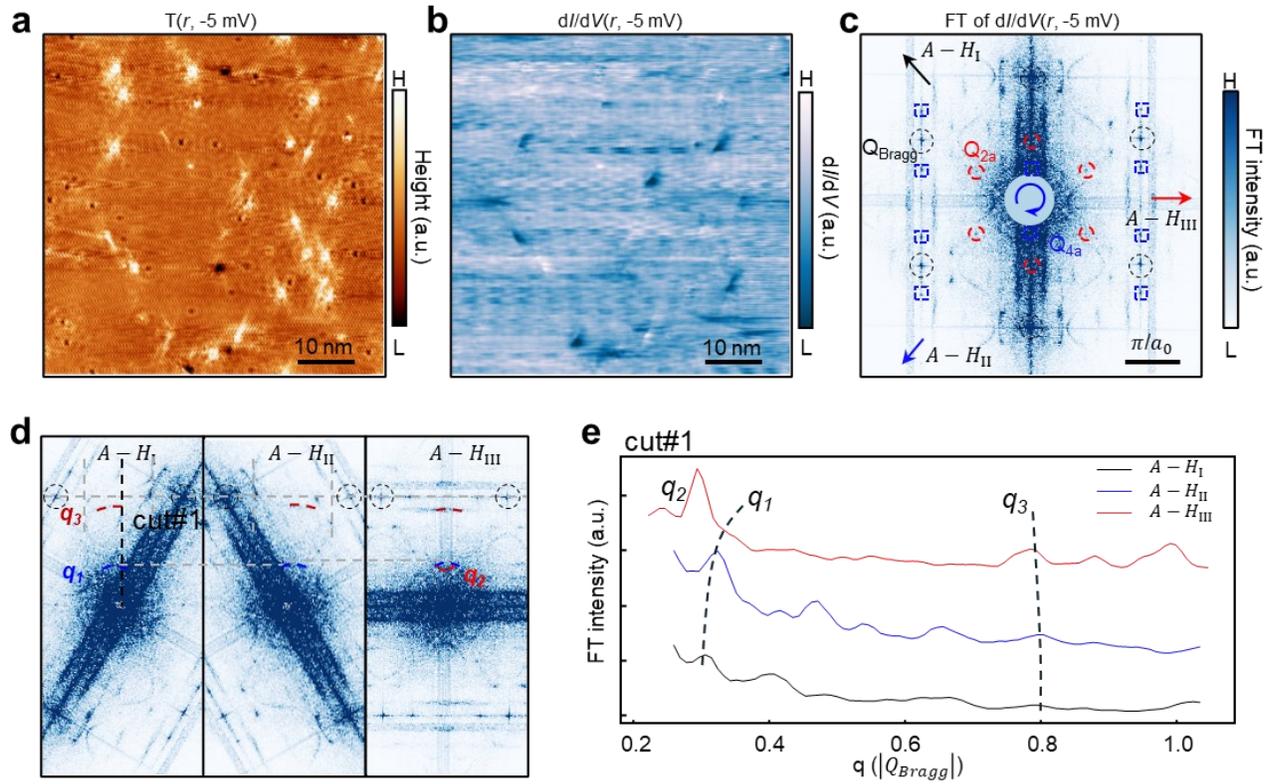

**Extended Data Fig. 3. Signature of chiral QPI in pristine CsV$_3$Sb$_5$ sample #1. a**, STM topographic image of pristine CsV$_3$Sb$_5$. $V_{bias}$= -5 mV, $I$= 500 pA. **b**, d$I$/d$V$ map of (**a**). $V_{bias}$= -5 mV, $I$= 500 pA, $V_{lock\text{-}in}$= 0.25 mV. **c**, FT of (**b**). Black dotted circles, red dotted circles, and blue dotted rectangles mark the scattering vectors of Bragg points, $2a_0 \times 2a_0$ CDW points, and $1a_0 \times 4a_0$ stripes points, respectively. $A-H_{\text{I}}$, $A-H_{\text{II}}$, and $A-H_{\text{III}}$ directions are indicated by black, blue, and red arrows, respectively. The curved arrow denotes the chirality of QPI patterns. **d**, The QPI patterns along $A-H_{\text{I}}$, $A-H_{\text{II}}$, and $A-H_{\text{III}}$ directions, highlighting the difference in three directions and showing the mirror symmetry breaking of the QPI patterns. **e**, Line profiles (cut#1) along $A-H_{\text{I}}$, $A-H_{\text{II}}$, and $A-H_{\text{III}}$ directions in (**d**), showing the different lengths of $q_1$ and $q_3$. The chirality of $q_1$ and $q_3$ are indicated by dotted curves.

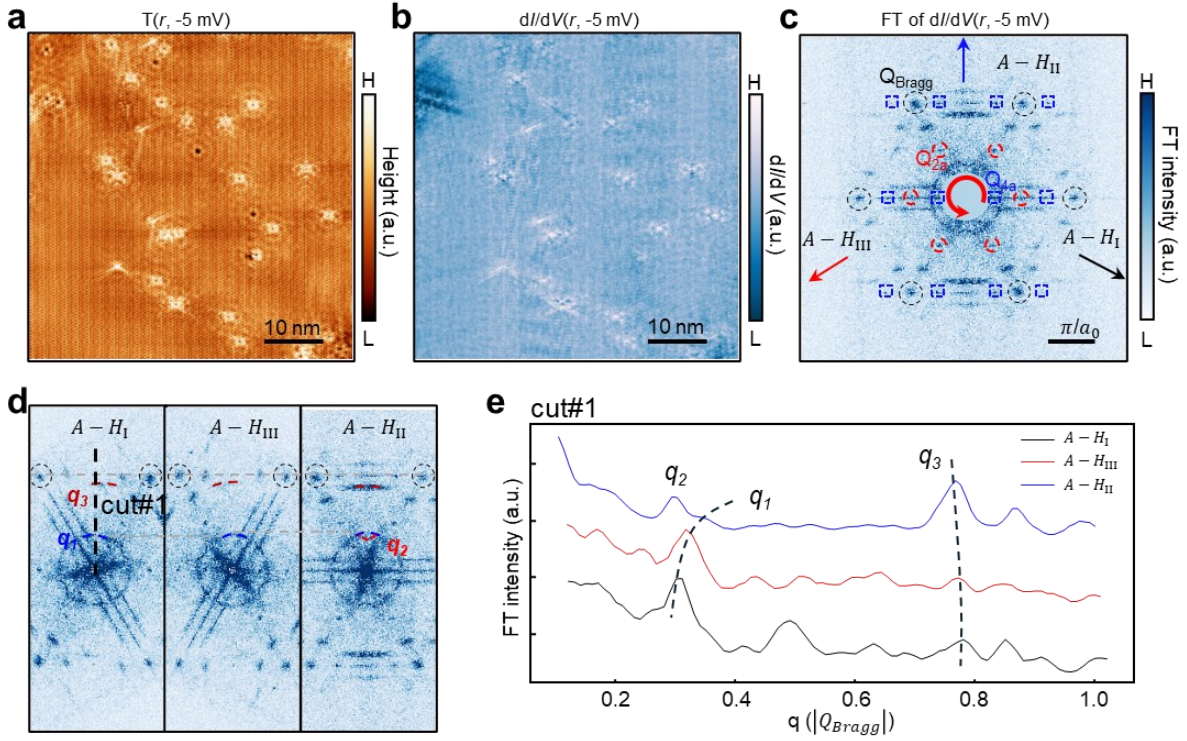

**Extended Data Fig. 4. Signature of chiral QPI in pristine CsV$_3$Sb$_5$ sample #2**. **a**, STM topographic image of pristine CsV$_3$Sb$_5$. $V_{bias}$= -5 mV, $I$= 500 pA. **b**, d$I$/d$V$ map of (**a**). $V_{bias}$= -5 mV, $I$= 500 pA, $V_{lock-in}$= 0.25 mV. **c**, FT of (**b**). Black dotted circles, red dotted circles, and blue dotted rectangles mark the scattering vectors of Bragg points, $2a_0 \times 2a_0$ CDW points, and $1a_0 \times 4a_0$ stripes points, respectively. $A - H_I$, $A - H_{II}$, and $A - H_{III}$ directions are indicated by arrows. The curved arrow denotes the chirality of QPI patterns. **d**, The QPI along $A - H_I$, $A - H_{II}$, and $A - H_{III}$ directions, highlighting the difference in three directions and showing the mirror symmetry breaking of the QPI patterns. **e**, Line profiles (cut#1) along $A - H_I$, $A - H_{II}$, and $A - H_{III}$ directions in (**d**). showing the different lengths of $q_1$ and $q_3$. The chirality of $q_1$ and $q_3$ are indicated by dotted curves.

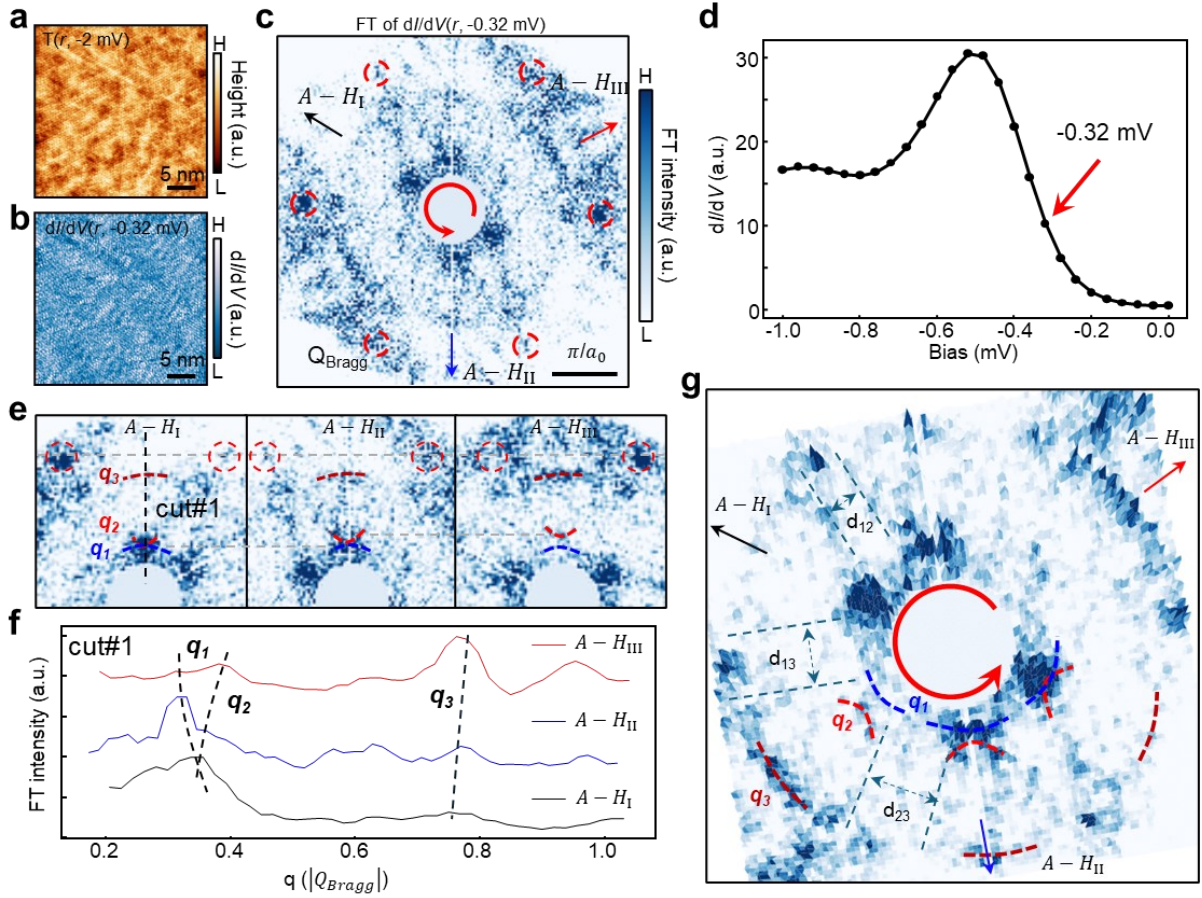

**Extended Data Fig. 5. Bogoliubov chiral QPI in CsV$_{3-x}$Ti$_x$Sb$_5$ ($x$=0.15).** **a**, STM topographic image of CsV$_{3-x}$Ti$_x$Sb$_5$ ($x$=0.15). $V_{bias}$= -2 mV, $I$= 500 pA. **b**, d$I$/d$V$ map at in-gap bias of -0.32 mV (indicated by arrow in (**d**)) of (**a**). $V_{bias}$= -2 mV, $I$= 500 pA, $V_{lock-in}$= 0.1 mV. **c**, FT of (**b**). Red dotted circles mark the Bragg scattering vectors. $A-H_{\mathrm{I}}$, $A-H_{\mathrm{II}}$, and $A-H_{\mathrm{III}}$ directions are indicated by the arrows. The curved arrow denotes the observed chirality. **d**, Spatially averaged d$I$/d$V$ spectrum on (**a**). $V_{bias}$= -2 mV, $I$= 500 pA, $V_{lock-in}$= 0.1 mV. **e**, The QPI patterns along $A-H_{\mathrm{I}}$, $A-H_{\mathrm{II}}$, and $A-H_{\mathrm{III}}$ directions, highlighting the difference in three directions and showing the mirror symmetry breaking of the QPI patterns. **f**, Line profiles (cut#1) along $A-H_{\mathrm{I}}$, $A-H_{\mathrm{II}}$, and $A-H_{\mathrm{III}}$ directions extracted from (**c**), showing the different lengths of scattering vectors **q$_1$**, **q$_2$**, and **q$_3$**, respectively, as indicated by dotted curves. **g**, 3D plot of the zoom-in FT in (**c**), highlighting the QPI patterns. The curved arrow indicates the chirality of QPI patterns. The central region is masked for clarity.

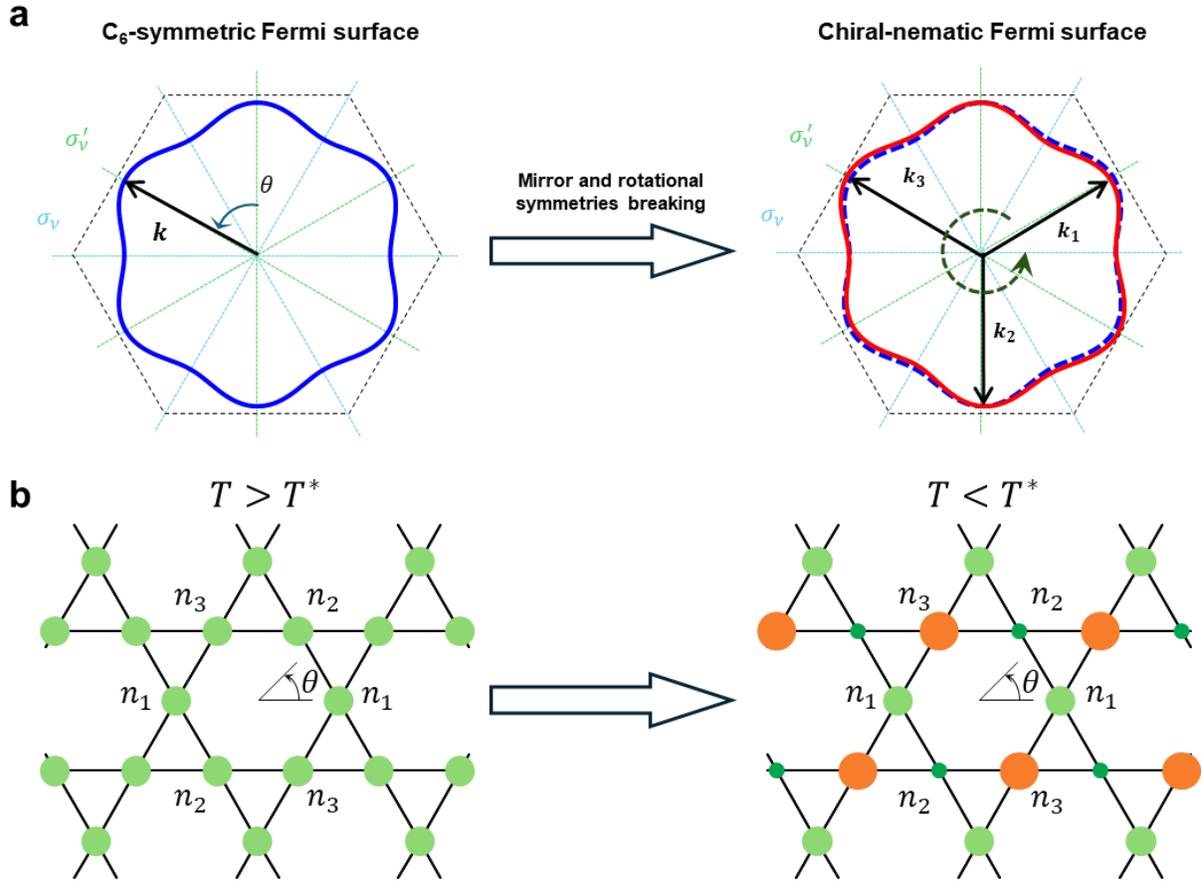

**Extended Data Fig. 6. Possible origin of the chiral-nematic Fermi liquid state. a,** Schematic of a chiral FS breaking all mirror symmetries. A general FS with $C_6$ rotational symmetry is plotted as (left panel) $k_F^0(\theta) = a \cdot \cos(6\theta) + b$ inside a hexagonal Brillouin zone. Two sets of mirror plane perpendicular to the kagome plane are indicated by $\sigma_v$ and $\sigma_v'$. A $d$-wave Pomeranchuk distortion of the symmetric FS according to $k_F(\theta) = k_F^0(\theta) + \eta \cdot \cos(2\theta - \pi/2)$ produces the chiral FS that breaks all mirror reflections $\sigma_v$ and $\sigma_v'$ (right panel). The chirality or handedness can be defined by the progressively decreasing amplitudes of the Fermi wave vectors $k_{3,2,1}$ along the crystal directions in the right panel. **b,** Schematic of a kagome lattice exhibiting chiral-nematic charge order. Within each unit cell, the onsite charge density varies among the three inequivalent sites, illustrated by the atom sizes. The charge density distribution is symmetric with $n_1 = n_2 = n_3$ for $T > T^*$. Below $T^*$, for generic $\varphi \neq 0, \pm\pi$, all mirror symmetries are broken and the pattern becomes chiral-nematic with $n_1 \neq n_2 \neq n_3$.